\documentclass{article}
\usepackage{spconf,amsmath,graphicx,amsfonts,epsfig}
\usepackage{times}
\usepackage{amsbsy}
\usepackage{latexsym}
\usepackage{amssymb}

\title{Distributed Estimation for Adaptive Networks Based on \\ Serial-Inspired Diffusion }

\name{Cornelius T. Healy and Rodrigo C. de Lamare}
\address{CETUC - Pontifical Catholic University of Rio de Janeiro - PUC-RJ\\
Rio de Janeiro, Brazil\\
Email: cthealy@cetuc.puc-rio.br, delamare@cetuc.puc-rio.br}

\begin{document}
\ninept

\linespread{0.975}
\maketitle
\begin{abstract}
Distributed estimation and processing in networks modeled by graphs have received a great deal of interest recently, due to the benefits of decentralised processing in terms of performance and robustness to communications link failure between nodes of the network. Diffusion-based algorithms have been demonstrated to be among the most effective for distributed signal processing problems, through the combination of local node estimate updates and sharing of information with neighbour nodes through diffusion. In this work, we develop a serial-inspired approach based on message-passing strategies that provides a significant improvement in performance over prior art. The concept of serial processing in the graph has been successfully applied in sum-product based algorithms and here provides inspiration for an algorithm which makes use of the most up-to-date information in the graph in combination with the diffusion approach to offer improved performance.
\end{abstract}
\begin{keywords}
Diffusion networks, wireless sensor
networks, distributed processing.
\end{keywords}
\section{Introduction}
\label{sec:intro}

Distributed signal processing is an important tool for problems
which may be modeled by a graph of nodes working to estimate a
parameter of interest as it allows computations to be carried
locally at the individual nodes, avoiding the need for a centralised
processing unit and thus offering robustness to scenarios where the
communication links to that central node are subject to channel
effects. To achieve this each node makes use of its local
observations in combination with the estimates produced at neighbour
nodes to produce an improved estimate of the parameters of interest.
This improved estimate is then shared with all neighbours of the
node, leading to propagation of the information through the network.

The distributed estimation problem has been considered in terms of
incremental \cite{Bertsekas_incremental1}, consensus
\cite{Tsitsiklis_concensus1} and diffusion \cite{Lopes_diff1,
Cattivelli_diff2, Chen_diff3} strategies. The incremental strategy
in general demands a computationally costly operation for
identifying a path through the graph nodes upon which to operate,
while the diffusion based strategies have been demonstrated to be
superior to those based on consensus in terms of convergence,
performance and stability \cite{Tu_diff_concensus}.

Recent work on the diffusion strategies has included sparsity-aware
approaches of \cite{Liu_sparse_diff,
Chouvardas_sparse_diff,de2009adaptive,fa2010reduced,zhaocheng2012l1,Songcen_sparse1,damdc},
which exploit the knowledge that the parameter vector to be
estimated may be sparse. Related work on improving the combiners in
the information diffusion stage of the algorithm has been reported
in \cite{Takahashi_combs, Tu_combs}. The effect of the network
topology on the diffusion strategies and how it may be exploited has
been considered in
\cite{Lopes_topology_diff,clarke2012transmit,peng2016adaptive,Songcen_topology_diff,
Songcen_sparse2}. A number of studies have considered imperfect
communications links between nodes in the network
\cite{Zhao_noisy_diff, Rastegarnia_noisy_diff}. Significant work has
also been carried out on analysis of the diffusion strategies
\cite{Zhao_analysis_diff, Chen_analysis_diff}.

In this paper, the schedule of node estimate updates is considered
as a source of performance improvement. This is motivated by the
observation that such an approach in the case of the sum-product
algorithm operating on a bipartite graph, as for the decoding of
LDPC codes, offers significantly faster convergence at almost no
cost in terms of additional complexity. This approach was termed
serial, shuffled or layered scheduling in the literature
\cite{Zhang_shuffledBP, Hocevar_LBP}. Further improvements were
found through more advanced update schedules
\cite{Healy1,healy2016design,Healy2,de2008minimum,de2013adaptive}.
Effectively, after each individual node update, the newly updated
messages are made available to the neighbour nodes of the updated
node, ensuring those neighbours compute their own updates with more
up-to-date and accurate information. As the diffusion approach is
based on the sharing of information with neighbour nodes in the
graph of the network, this concept of serialisation translates well.
In particular, we develop a serial-inspired (SI) least-mean square
(LMS) , which we denote SI-LMS and can exploit the schedule of node
updates to  obtain improved performance. In the proposed SI-LMS
algorithm, serialisation is introduced through the inclusion of an
additional diffusion combination which has access to the most
recently updated estimates in the graph. The proposed SI-LMS
algorithm offers significant improvements in convergence speed, as
is demonstrated by the simulation study provided in this paper.

This paper is organized as follows. Section \ref{sec:prob} provides
the problem statement and introduces the diffusion strategy for
system identification using the LMS algorithm. In Section
\ref{sec:prop_alg} the proposed algorithm is developed and described
in detail, along with pseudocode representation. Section
\ref{sec:results} provides the numerical simulation results, and
Section \ref{sec:concl} concludes the paper.

\textbf{Notation:} Throughout this paper, lowercase letters such as
$x$ indicate scalars, lowercase boldface letters such as
${\boldsymbol x}$ denote column vectors and uppercase boldface
letters such as ${\boldsymbol A}$ denote matrices. The superscript
$i$ denotes that ${\boldsymbol A}^{(i)}$ is the realization of
${\boldsymbol A}$ at the time index $i$, likewise for scalars as in
$d^{(i)}$. Subscripts are used to identify the node or nodes in the
graph with which a value is associated.


\section{Problem Statement and the Diffusion Strategy}
\label{sec:prob}

\begin{figure}[ht!]
\centering
\includegraphics[width=0.3\columnwidth]{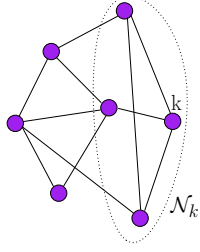}
\caption{An example of a connected graph with neighbourhood $\mathcal{N}_k$ of node $k$ shown.}
\label{fig:network_diag}
\end{figure}

Consider a network modeled by a graph with N nodes as depicted in
Fig. \ref{fig:network_diag}. The value $d_k^{(i)}$ is the scalar
observation at time instant $i$ for the node $k$ in the graph, and
the observation is related to the input signal ${{\boldsymbol
x}_k^{(i)}}$ by
\begin{equation}
\label{eqn:syst_model}
{d_k^{(i)}} = {{\boldsymbol \omega}}_0^H{\boldsymbol x_k^{(i)}} +{n_k^{(i)}},~~~
i=1,2, \ldots, N ,
\end{equation}
where the input signal vector ${{\boldsymbol x}_k^{(i)}}$ for node $k$ at time index $i$ is an $M \times 1$ vector. The value ${ n_k^{(i)}}$ is the noise sample at node $k$ and time index $i$ and has zero mean and variance $\sigma_{v,k}^2$. The goal of the distributed estimation problem is to estimate the value of ${\boldsymbol \omega}_0$ based on the knowledge at the nodes in the network of the observations $d_k^{(i)}$, the input signal vectors ${{\boldsymbol x}_k^{(i)}}$ and the relation in (\ref{eqn:syst_model}) through use of that local knowledge and the ability to share information with neighbours in the network graph.


The diffusion strategy for distributed estimation involves a process of local adaptation with the information available using for example the LMS estimate update, followed by information sharing with neighbour nodes involving a weighted sum of the estimates across the neighbourhood of each node. This process leads to diffusion of information through the fully connected graph. The adaptation and combination steps of the diffusion strategy can be performed in either order, leading to in one case the adapt--then--combine (ATC) diffusion strategy and in the other the  combine--then--adapt (CTA) diffusion strategy \cite{Lopes_diff1}. The two steps of ATC diffusion are described by

\begin{equation}
\label{eqn:ATC_diff1}
{\boldsymbol \psi}_k^{(i)}= {\boldsymbol \omega}_k^{(i-1)}+{\mu}_k {{\boldsymbol x}_k^{(i)}}[{ d_k^{(i)}}-
{\boldsymbol \omega}_k^{(i-1)H}{\boldsymbol x_k^{(i)}}]^*,
\end{equation}
\begin{equation}
\label{eqn:ATC_diff2}
{\boldsymbol \omega}_k^{(i)}= \sum\limits_{l\in \mathcal{N}_k} c_{kl} {\boldsymbol \psi}_l^{(i)},
\end{equation}
where the values $c_{kl}$ are known as the combination coefficients
and provide the weighting in the combination step of the respective
algorithms and ${\boldsymbol w}_k$ represents the parameter
estimator. They are related to the topology of the graph, being
nonzero only if node $k$ and node $l$ are neighbours and
additionally must satisfy the constraint:
\begin{equation}
\sum\limits_{l} c_{kl} =1 , l\in \mathcal{N}_k \forall k .
\end{equation}
There are a number of rules specifying the combination coefficients
to be found in the literature, including the uniform , Metropolis ,
relative degree and Laplacian rules. The Metropolis combiner is
given by:
\begin{equation}
\left\{\begin{array}{ll}
c_{kl}= \frac{1}{\max(|\mathcal{N}_l|,|\mathcal{N}_k|)},~~~~~~~\rm{if}~l \in \mathcal{N}_k,~l \neq k\\
c_{kl}=  1 - \sum_{l \in \mathcal{N}_k/k}{c_{kl}},~~~\rm{if}~l \in \mathcal{N}_k,~l = k\\
c_{kl}=0,~~~\rm{if}~ l \notin \mathcal{N}_k\\
\end{array}
\right.
\end{equation}

\begin{figure}[ht!]
\centering
\includegraphics[width=0.9\columnwidth]{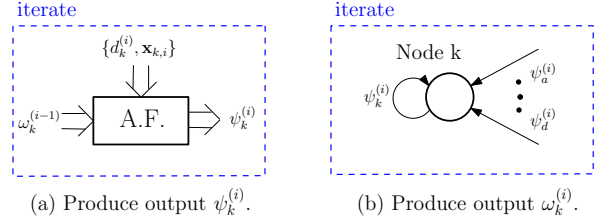}
\caption{(a) Filter adaptation with local observations and combined
estimate. (b) Information diffusion through weighted combination of
neighbour estimates.} \label{atc_blockdiag}
\end{figure}

%
Fig. \ref{atc_blockdiag} provides the block diagram for the ATC
diffusion algorithm which implements (\ref{eqn:ATC_diff1}) -
(\ref{eqn:ATC_diff2}).

\section{Proposed Serial-inspired algorithm}
\label{sec:prop_alg}

In this section, the proposed SI-LMS algorithm is introduced. In the
case of sum-product type algorithms operating on bipartite graphs,
the standard message update schedule is to activate all nodes of one
type and then to activate all nodes of the other type. It was
demonstrated that improvements in error rate convergence of the
algorithms may be found at no increased computational complexity if
the nodes in the graph are updated in a serial fashion, also termed
shuffled or layered schedule \cite{Zhang_shuffledBP, Hocevar_LBP}.
This improvement in convergence behaviour is derived from the fact
that the most recently updated messages in the graph may be used to
improve the next updates within an iteration. This observation has
motivated the investigation of the diffusion LMS algorithm and
resulted in the development of the algorithm proposed in this paper.
Essentially, the process of information diffusion is used to improve
the LMS adaptation through the use of new estimates at neighbour
nodes as soon as they are available.

The MSE cost function at node $k$ takes the form
\begin{equation}
\label{eqn:MSE_cost}
J_k({\boldsymbol\omega}) = \mathbb{E}[|d_k^{(i)} - {\boldsymbol\omega}^H{\boldsymbol x_k^{(i)}}|^2],
\end{equation}
which through expansion and rearrangement results in
\begin{equation}
J_k({\boldsymbol\omega}) = J_{k,min} + ||{\boldsymbol\omega} - {\boldsymbol\omega}_0||_{{\boldsymbol R}_{x,k}}^2,
\end{equation}
where $J_{k,min}$ is the value of $J_k({\boldsymbol\omega})$
evaluated at ${\boldsymbol\omega} = {\boldsymbol\omega}_0$. The
local cost function at node $k$ when information sharing with
neighbours is allowed becomes:
\begin{equation}
J_k^{local}({\boldsymbol\omega}) = \sum_{l \in \mathcal{N}_k} a_{l,k} J_l({\boldsymbol\omega}).
\end{equation}
The global cost function is simply the sum across all nodes $l$ of
this local cost function. Rewritten in terms of the node of interest
and its neighbours, we have the global function from the perspective
of node $k$:
\begin{equation}
J_k^{global}({\boldsymbol\omega}) = J_k^{local}({\boldsymbol\omega})  + \sum_{l \in \mathcal{N}_k\backslash \{k\}} J_l^{local}({\boldsymbol\omega}).
\end{equation}
Applying the steepest-descent method we arrive at the recursion for
updating the parameters of the estimator:
\begin{equation}
{\boldsymbol\omega}_k^{(i)} = {\boldsymbol\omega}_k^{(i-1)} - \mu_k[\nabla_{\boldsymbol\omega} J_k^{global}({\boldsymbol\omega}_k^{(i-1)})]^*,
\end{equation}
where $\nabla_{\boldsymbol\omega}
J_k^{global}({\boldsymbol\omega}_k^{(i-1)})$ is the gradient of
$J_k^{global}({\boldsymbol\omega})$ with respect to
${\boldsymbol\omega}$ evaluated at ${\boldsymbol\omega}_k^{(i-1)}$.
Through the use of the expanded versions of (\ref{eqn:MSE_cost}) and
through a number of approximations detailed in
\cite{Cattivelli_diff2, Sayed_book_chapter}, the update recursion
may be reformulated as:
\begin{multline}
\label{eqn:est_upd}
{\boldsymbol\omega}_k^{(i)} = {\boldsymbol\omega}_k^{(i-1)} + \mu_k\sum_{l \in \mathcal{N}_k} a_{l,k}  ({\boldsymbol r}_{dx,l} - {\boldsymbol R}_{x,l}{\boldsymbol\omega}_k^{(i-1)}) \\ +  \mu_k\sum_{l \in \mathcal{N}_k\backslash \{k\}} b_{l,k} ({\boldsymbol\omega}_0 - {\boldsymbol\omega}_k^{(i-1)}),
\end{multline}
where ${\boldsymbol R}_{x,l} = \mathbb{E} [{\boldsymbol x}_k^{(i)}
{\boldsymbol x}_k^{(i)H}]$ and ${\boldsymbol r}_{dx,l} = \mathbb{E}
[d_k^{(i)*} {\boldsymbol x}_k^{(i)}]$. In (\ref{eqn:est_upd}) the
previous estimate is corrected by a filter adaptation term and an
information diffusion term. In the development of the ATC and CTA
algorithms, the two correction terms are applied successively. In
the proposed algorithm, the information diffusion term will be
applied both before and after the adaptation of the estimator, and
as in the development of those algorithms the best available
estimate will be used to substitute for both ${\boldsymbol\omega}_0$
and ${\boldsymbol\omega}_k^{(i-1)}$ in (\ref{eqn:est_upd}). In the
proposed algorithm these best estimates are improved upon the
previously presented works through the observation that the
estimates which have been updated at neighbour nodes are available
for use immediately, and so the first diffusion correction term is
applied in a serial fashion. In particular, we employ instantaneous
estimates of ${\boldsymbol R}_{x,l}$ and ${\boldsymbol r}_{dx,l}$ to
obtain the recursion for the proposed SI-LMS algorithm:
\begin{multline}
\label{eqn:est_upd_inst} {\boldsymbol\omega}_k^{(i)} =
{\boldsymbol\omega}_k^{(i-1)} + \mu_k\sum_{l \in \mathcal{N}_k}
a_{l,k}  {\boldsymbol x}_k^{(i)}[d_k^{(i)} -
{\boldsymbol\omega}_k^{(i-1)^{H}}{\boldsymbol x}_k^{(i)}]^* \\ +
\mu_k\sum_{l \in \mathcal{N}_k\backslash \{k\}} b_{l,k}
({\boldsymbol\omega}_0 - {\boldsymbol\omega}_k^{(i-1)}),
\end{multline}
Given the previous development, the steps of the proposed SI-LMS
algorithm are as follows:
\begin{enumerate}
\item Combine prior and new estimates available from neighbourhood nodes, including the node of interest, by weighted sum.
\item Adapt the parameters according to the chosen rule, using local observations at the node and combined estimate from the first step. Make new estimate available to neighbour nodes.
\item Combine estimates available from neighbourhood.
\end{enumerate}

The SI-LMS algorithm is presented as
\begin{equation}
\label{eqn:CTATC_diff1}
\displaystyle {\boldsymbol \psi}_k^{(i)} = c_{k,k}{\boldsymbol \omega}_k^{(i-1)} + \sum_{\ell \in \mathcal{N}_k: \ell \geq k}{c_{k,\ell}{\boldsymbol \omega}_\ell^{(i-1)}} + \sum_{m \in \mathcal{N}_k: m < k}{c_{k,m}{\boldsymbol \phi}_m^{(i)}},
\end{equation}
\begin{equation}
\label{eqn:CTATC_diff2} \displaystyle {\boldsymbol \phi}_k^{(i)} =
{\boldsymbol \psi}_k^{(i)} + \mu_k {\boldsymbol x}_k^{(i)}[d_k^{(i)}
- {\boldsymbol \psi}_k^{(i)^{H}} {\boldsymbol x}_k^{(i)}]^{*},
\end{equation}
\begin{equation}
\label{eqn:CTATC_diff3}
\displaystyle {\boldsymbol \omega}_p^{(i)} = \sum_{q \in \mathcal{N}_p}{a_{p,q}{\boldsymbol \phi}_q^{(i)}},
\end{equation}

\begin{figure}[ht!]
\centering
\includegraphics[width=\columnwidth]{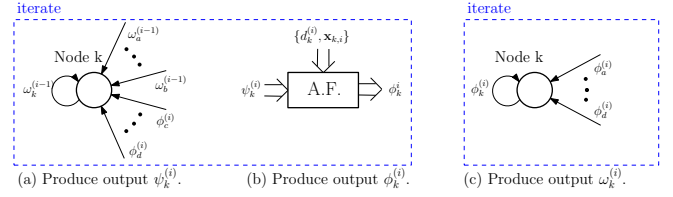}
\caption{(a)Weighted sum of updated and prior estimates at the
neighbours. (b) Filter adaptation with local observations and
combined estimate. (c) Final weighted combination of neighbour
estimates.} \label{block_diag}
\end{figure}

Fig. \ref{block_diag} provides a block diagram of the proposed
SI-LMS algorithm, with blocks for the initial serial information
diffusion, the adaptation for the estimator and the final diffusion
combination, respectively. In Alg. \ref{alg:CTATC_LMS} the
pseudocode for the proposed SI-LMS algorithm is provided. This gives
the details of the algorithm.

\subsection{Computational Cost and Bandwidth Requirements of the Proposed SI-LMS Algorithm}
\label{subsec:compl_comm_costs}

The proposed SI-LMS algorithm provides improvements in convergence speed through the use of an additional combination of new and old estimates prior to the LMS estimate adaptation. This additional weighted sum prior to the adaptation step (line 4 in Algorithm \ref{alg:CTATC_LMS}) comprises the only additional cost of the proposed algorithm. Thus, in terms of complexity per node, the proposed algorithm costs

\begin{itemize}
\item $|\mathcal{N}_k|$ extra multiplications
\item $|\mathcal{N}_k|$ extra additions
\end{itemize}

With the complexity cost across the network being $N$ times each of these. Note that in addition to the cost in terms of increased complexity, the proposed SI-LMS algorithm requires that the nodes in the graph share their updated estimates with their neighbours as soon as they are produced by the LMS adaptation. Thus the SI-LMS algorithm also incurs an extra communications cost of

\begin{itemize}
\item $N(|\mathcal{N}_k|-1)$ transmissions of a vector of dimension $M \times 1$
\end{itemize}

when compared to the ATC-LMS diffusion algorithm \cite{Lopes_diff1}.

\section{Simulation Results}
\label{sec:results}

\begin{figure}[ht!]
\centering
\includegraphics[width=.85\columnwidth]{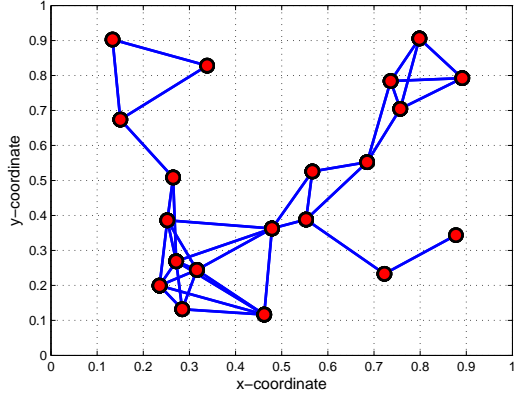}
\caption{The topology of the network for the results of Figs. \ref{res:node_powers} to \ref{res:MSE_diffpower}.}
\label{res:network_topology}
\end{figure}

\begin{figure}[ht!]
\centering
\includegraphics[width=0.9\columnwidth]{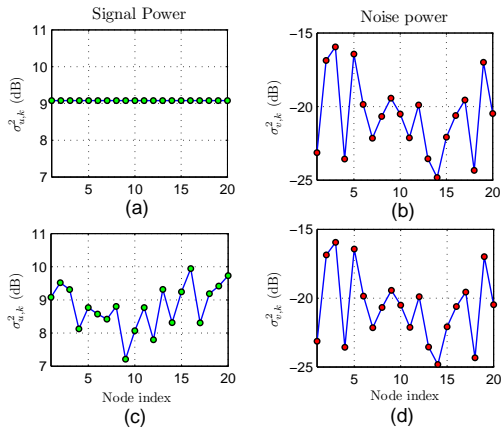}
\caption{The details of signal and noise power at the nodes in the
network of Fig. \ref{res:network_topology} considered in the results
of Figs. \ref{res:MSE_samepower} and \ref{res:MSE_diffpower},
respectively.} \label{res:node_powers}
\end{figure}

\begin{figure}[ht!]
\centering
\includegraphics[width=0.85\columnwidth]{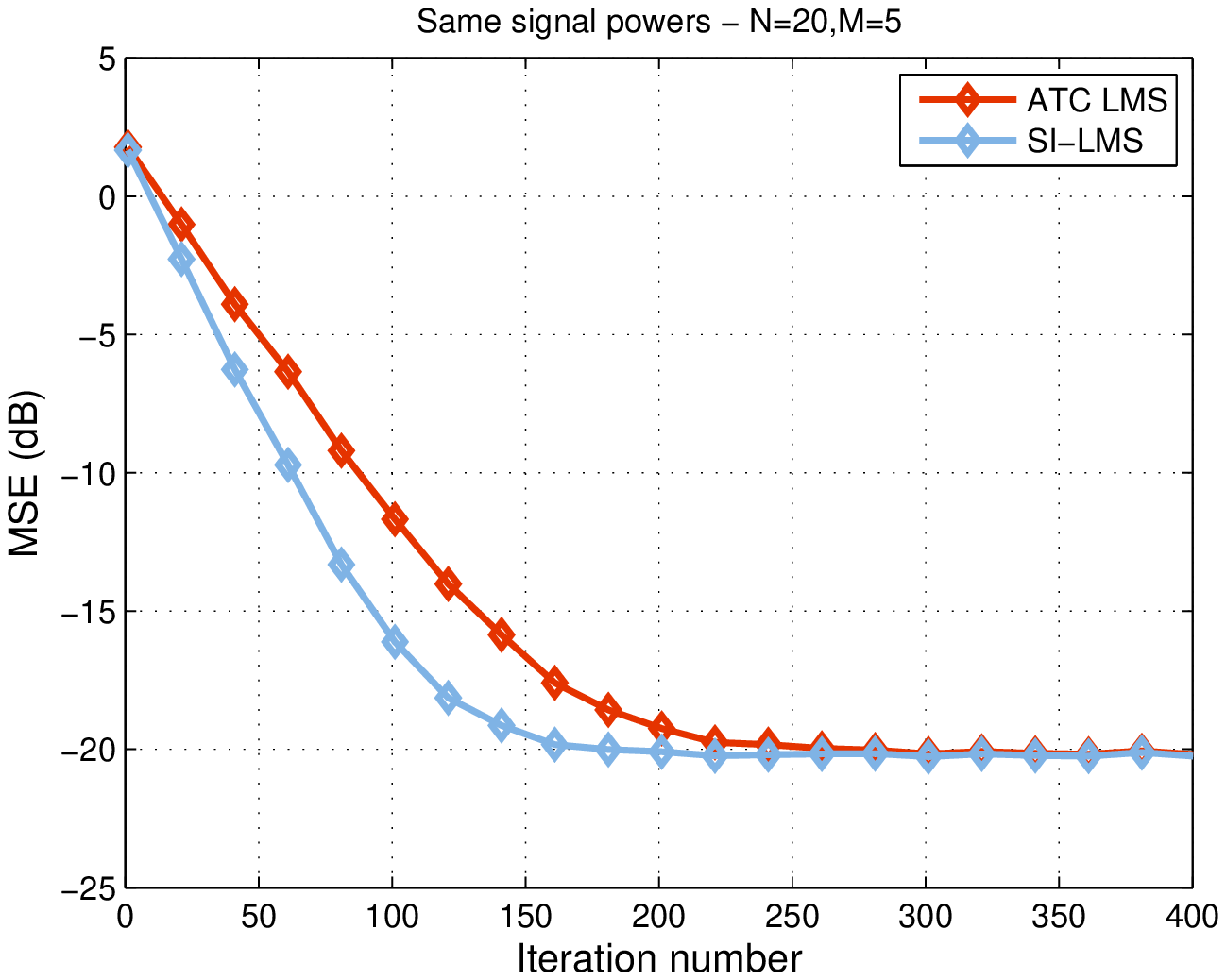}
\caption{The network MSE of the network with topology of Fig.
\ref{res:network_topology} and the signal and noise variance
parameters provided in of Fig. \ref{res:node_powers}(a) and
\ref{res:node_powers}(b).} \label{res:MSE_samepower}
\end{figure}

\begin{figure}[ht!]
\centering
\includegraphics[width=0.85\columnwidth]{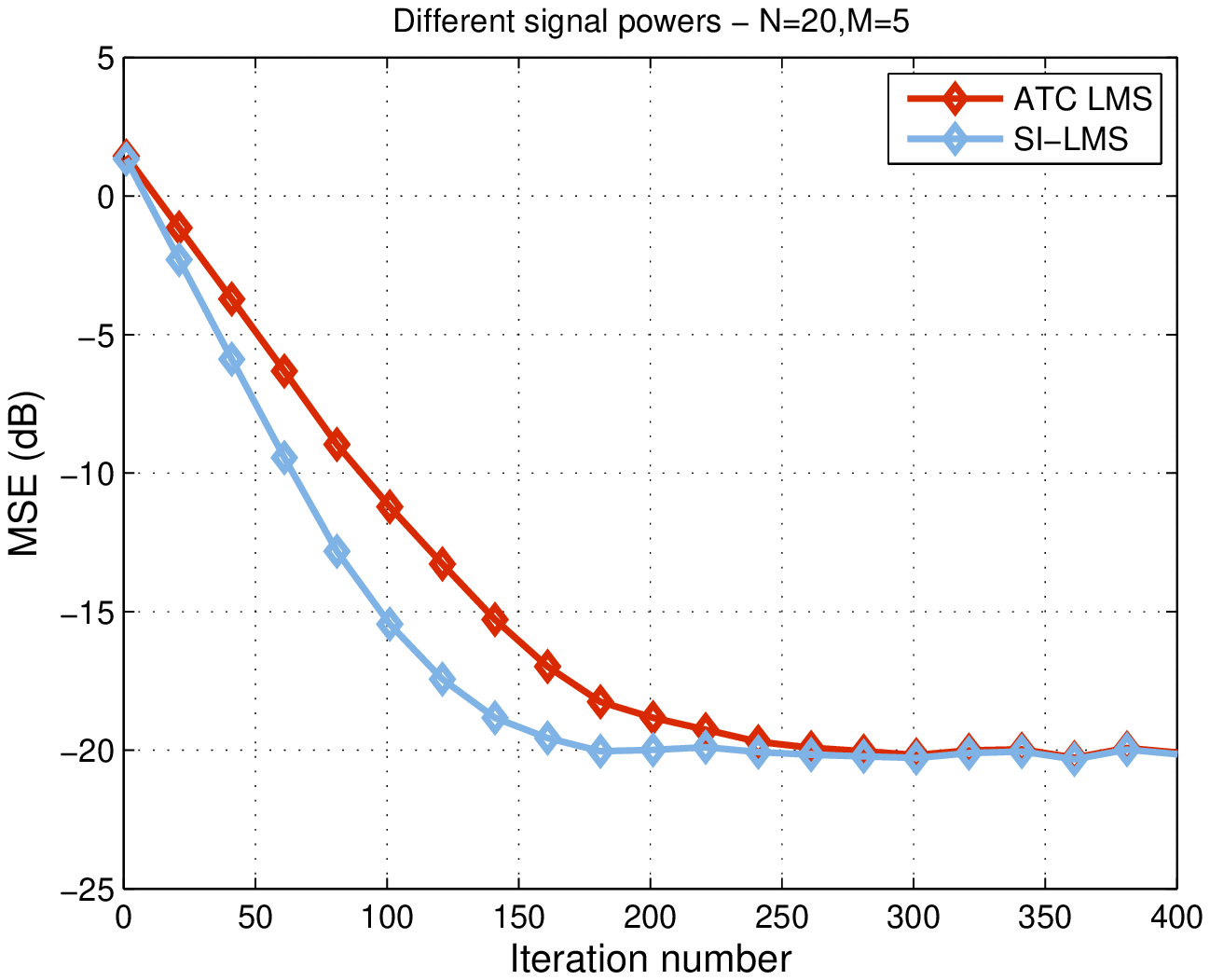}
\caption{The network MSE of the network with topology of Fig.
\ref{res:network_topology} and the signal and noise variance
parameters provided in of Fig. \ref{res:node_powers}(c) and
\ref{res:node_powers}(d).} \label{res:MSE_diffpower}
\end{figure}

\begin{figure}[ht!]
\centering
\includegraphics[width=0.85\columnwidth]{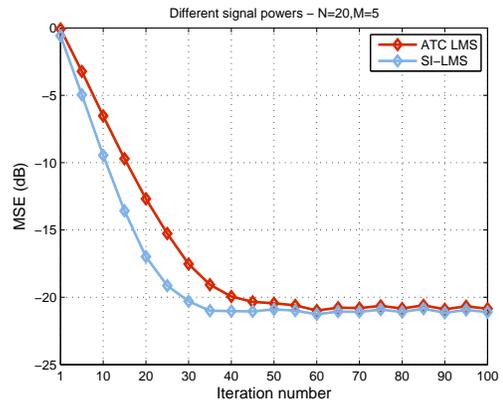}
\caption{The network MSE of the network with topology of Fig.
\ref{res:network_topology} and the variance parameters provided in
of Fig. \ref{res:node_powers}(c) and \ref{res:node_powers}(d) for
the case when a larger step size is used in the adaptive
algorithms.} \label{res:MSE_diffpower_largermu}
\end{figure}

In this section, the simulation study for the proposed algorithm is
presented. Its performance, in terms of mean-square error (MSE), is
compared to the diffusion ATC algorithm \cite{Lopes_diff1}. Fig.
\ref{res:network_topology} provides the network graph topology,
showing the network considered has $N=20$ nodes. We adopted the
Metropolis combining rule \cite{Sayed_book_chapter}. The unknown
parameter vector to be estimated has length $M=5$. Two cases are
considered for the input signal, one with the signal variance equal
at all nodes in the network, another with varying signal variances.
The noise of (\ref{eqn:syst_model}) is modeled by white complex
circular Gaussian random variables with zero mean, with signal
variances that are arbitrary. Two cases are considered, the first in
which the signal variances are the same across all nodes in the
network, and another where they are allowed to vary. The variances
are provided in Fig. \ref{res:node_powers}. The variances of Fig.
\ref{res:node_powers}(a) and \ref{res:node_powers}(b) correspond to
the simulation environment for the results of Fig.
\ref{res:MSE_samepower} while the variances of Fig.
\ref{res:node_powers}(c) and \ref{res:node_powers}(d) correspond to
the simulation environment for the results of Fig.
\ref{res:MSE_diffpower}. The step size at all nodes in the network
is $\mu_k = 0.01$ for both Fig. \ref{res:MSE_samepower} and Fig.
\ref{res:MSE_diffpower}. Additional results for the scenario with
different variances at the nodes in the network are provided for a
larger step size of $\mu_k = 0.05$ in Fig.
\ref{res:MSE_diffpower_largermu}. The results provided are averaged
over $100$ independent runs.

Figs. \ref{res:MSE_samepower} and \ref{res:MSE_diffpower}
demonstrate that the proposed SI-LMS algorithm outperforms the
standard-form ACT-LMS algorithm in speed of convergence, with
approximately a $40\%$ reduction in the number of iterations
required to converge. Fig. \ref{res:MSE_diffpower_largermu} shows
that the performance improvements of the proposed SI-LMS algorithm
are consistent.

%
\section{Conclusion}
\label{sec:concl}

In this paper, a diffusion-based SI-LMS algorithm has been
presented, which exploited the most recent estimates available in
the network graph to improve the convergence of the estimates
throughout the network. This was achieved through the inclusion of
an additional information diffusion step, which is carried out in a
serial manner. A discussion of the costs of the proposed SI-LMS
algorithm in terms of increased computation required at the nodes in
the graph and additional necessary communication of estimates
required for information diffusion was provided, demonstrating that
the proposed algorithm is not prohibitively costly considering the
benefits offered. Numerical results justify the proposed SI-LMS
algorithm and illustrate its performance advantages.

\bibliographystyle{IEEEbib}
\bibliography{reference_ICASSP}

\end{document}